# Stabilization of the first-order phase transition character and Enhancement of the Electrocaloric Effect by NBT substitution in BaTiO₃ ceramics

Merve Karakaya[a], İrem Gürbüz[a,b], Lovro Fulanovic[c], Umut Adem[a,*]

Electrocaloric properties of BT- based Pb-free ferroelectric materials are widely investigated. One approach to achieve a large electrocaloric response is making use of the substantial polarization change associated with the first-order phase transition at the Curie temperature. To make use of that approach, we have investigated the electrocaloric response of (1-x)BaTiO₃-xNa₀.₅Bi₀.₅TiO₃ (BT-NBT) ceramics with x = 0.05, 0.10, 0.20 and 0.30. For this BT-rich part of the solid solution, it is established that increasing NBT content increases the tetragonality of the BaTiO₃. We show that this increase in tetragonality with NBT substitution helps to maintain the first-order nature of the phase transition in BaTiO₃ and correspondingly large electrocaloric response, despite the simultaneous enhancement of relaxor ferroelectric character with the NBT substitution. A significantly larger effective electrocaloric temperature change ($\Delta T_{eff}$) of 1.65 K was obtained for x = 0.20 sample under 40 kV/cm, using the direct measurement of the electrocaloric effect, which is in reasonable agreement with the indirect measurements.

## Introduction

As a lead-free piezoelectric, Na₀.₅Bi₀.₅TiO₃ (NBT) based materials have been widely studied, especially as a solid solution with BaTiO₃.[1-3] Enhanced piezoelectric and dielectric properties of (1-x) Na₀.₅Bi₀.₅TiO₃–xBaTiO₃ (NBT-BT) solid solution have been obtained at rhombohedral-tetragonal morphotropic phase boundary (MPB), which corresponds to x = 0.06–0.11 compositions.[4] However, few studies are focused on the BT-rich compositions, BaTiO₃ (1-x)BaTiO₃– xNa₀.₅Bi₀.₅TiO₃ (BT-NBT).[5-8] These show that incorporation of NBT into BT up to 40% causes an increase in the tetragonality ($c/a$) and Curie temperature (T_C). This is an unusual effect because apart from Pb²⁺, the substitution of ions at the A and B-sites of the perovskite structure decreases tetragonality and T_C.[9] The origin of the enhanced tetragonality was attributed to the decrease in the oxygen-octahedral volume rather than the off-centering of the B site Ti⁺⁴ ion.[6] Because the doping concentration above $x = 0.15$ makes a gradual decrease in $c$ and also more decrease in $a$ which effectively raises the tetragonality. Since BT-based materials are ubiquitously used as multilayer ceramic capacitors (MLCC), higher T_C of BT-rich BT-NBT solid solutions is exploited in MLCCs for high temperatures such as X8R and are also suitable for even higher temperatures, satisfying the criteria for X9R. The use of X7R and X8R type MLCCs as capacitors in environments with operating temperatures up to 200°C would not be suitable as their maximum temperatures are 125 °C and 150 °C, respectively.[10-12] Another area where high T_C BT-NBT based materials are exploited, is the positive temperature coefficient of resistivity (PTCR) materials that function at high temperatures.[13]

a. Department of Materials Science and Engineering, İzmir Institute of Technology, Urla, 35430, İzmir Turkey. E-mail:umutadem@iyte.edu.tr
b. Department of Chemistry, Aix-Marseille University, 52 Avenue Escadrille Normandie Niemen, 13013, Marseille, France.
c. Nonmetallic Inorganic Materials, Department of Materials and Earth Sciences, Technical University of Darmstadt, Peter-Grünberg-Straße 2, 64287 Darmstadt, Germany
† Electronic Supplementary Information (ESI) available

Order of the ferroelectric-paraelectric phase transition is critical for the electrocaloric effect. It is established that a 1$^{st}$ order phase transition with a sharper depolarization behaviour at T$_C$ leads to large adiabatic temperature change ($\Delta T$), albeit in a narrow temperature interval. Large $\Delta T$ around the Curie temperature due to the 1$^{st}$ order ferroelectric phase transition was reported for BaTiO$_3$ in single crystalline,[14] polycrystalline[15] and multilayer ceramic capacitor (MLCC)[16] forms. By substituting with Zr at the Ti-site, it is possible to introduce a diffuse, and depending on the amount of the substitution, even a relaxor ferroelectric character to the phase transition in BaTiO$_3$. The phase transitions in those cases are no longer first order. However, they take place over a broader temperature range which helps to extend the temperature range in which a relatively large electrocaloric response is sustained.[17] Generally, when B-site ion is substituted with ions of different charge and valence, due to the chemical disorder, relaxor ferroelectric character is developed in ferroelectric perovskite oxides[8]. One interesting exception is reported for the lead scandium tantalate Pb(Sc$_{0.5}$Ta$_{0.5}$)O$_3$, where the material keeps its normal ferroelectric behaviour with the first-order ferroelectric transition character in the cation-ordered state.[18] This leads to a large electrocaloric response, causing a $\Delta T$ of 3.7 K under an electric field of 40 kV/cm.[19]

In this study, we report on the phase transition character and the electrocaloric response of (1-x)BT-xNBT (x = 0.05, 0.1, 0.2 and 0.3) ceramics (BT-rich part of the NBT-BT solid solution) and we show that despite the increasing chemical disorder at the A-site, the ferroelectric character of BaTiO$_3$ is stabilized with increasing NBT content. The stabilization of ferroelectricity by NBT substitution causes the phase transition to remain first-order for x = 0.2 and x = 0.3, leading to a large electrocaloric response measured directly and indirectly.

## Experimental

The conventional solid-state reaction method was used to prepare the ceramic samples (1-x)BaTiO$_3$-xBi$_{0.5}$Na$_{0.5}$TiO$_3$ (BT-NBT) where x = 0.05, 0.10, 0.20 and 0.30. The high purity metal oxides and carbonates BaCO$_3$ (99%), Bi$_2$O$_3$ (99.9%), Na$_2$CO$_3$ (99.8%), and TiO$_2$ (99.9%) were used as starting raw materials. All powders were dried at 200 °C for overnight. Then they were weighed according to stoichiometric formula and mixed in a 30 ml HDPE bottle using zirconia balls in ethanol media by planetary ball milling for 18 h. The calcination process was carried out in pellet form at 1000 °C for 2 h. After the calcination, the pellets were ground into powder and mixed with 0.3 mol% MnO$_2$ (99%) powder to enhance the resistivity[20, 21]. The powders were mixed with 4 wt% polyvinyl alcohol (PVA) binder and water solution and then ball milled for 18 h. The resulting powder was dried, sieved and pressed into the disc form (~10 mm in diameter, ~1 mm thick) with a pressure ~100 MPa. Then the PVA binder was burned out at 600 °C for 4 h. After burnout the pellets were sintered at 1200 °C for 2 h with a heating rate of 5 °C/min. Sintering temperature of the samples was kept low due to the volatility of Na and Bi.

Powder X-ray diffraction (PXRD, Cu K$_\alpha$ radiation, Philips X'Pert Pro) was used to determine the crystal structures of the crushed sintered pellets. The microstructure of ceramics was studied on polished and thermally etched surfaces (thermal etching was done at 1100 °C, for 1 h) using a scanning electron microscope (SEM, FEI QUANTA 250 FEG). Differential scanning calorimetry (DSC) was used to monitor the phase transition character and to obtain the specific heat of the samples (with a Perkin Elmer, DSC 6000). For the electrical measurements, top and bottom surfaces of polished pellets were coated with silver paste. Ag paste was dried at 200 °C for 20 min.

The temperature dependence of dielectric properties of ceramics was measured at various frequencies (0.1, 1, 10, 100 kHz) by an LCR meter (KEYSIGHT, E4980AL). The polarization-electric field (P-E) hysteresis loops wereobtained by using AIXACCT TF Analyzer 1000 and a high voltage amplifier (TREK 610E). The field-induced strain data were also collected simultaneously by a laser-interferometer (SIOS). For all electrical property measurements, a sample holder (AIXACCT piezo sample holder TFA 423-7) was used and temperature was controlled by a PID controller. $\Delta T_{ad}$ is indirectly determined using $\Delta T = -\frac{1}{\rho}\int_{E_1}^{E_2}\frac{T}{C}\left(\frac{\partial P}{\partial T}\right)_E dE$, where the specific heat C is obtained from DSC measurements.[22] and the densities $\rho$ of samples that were measured by Archimedes' method have been used.

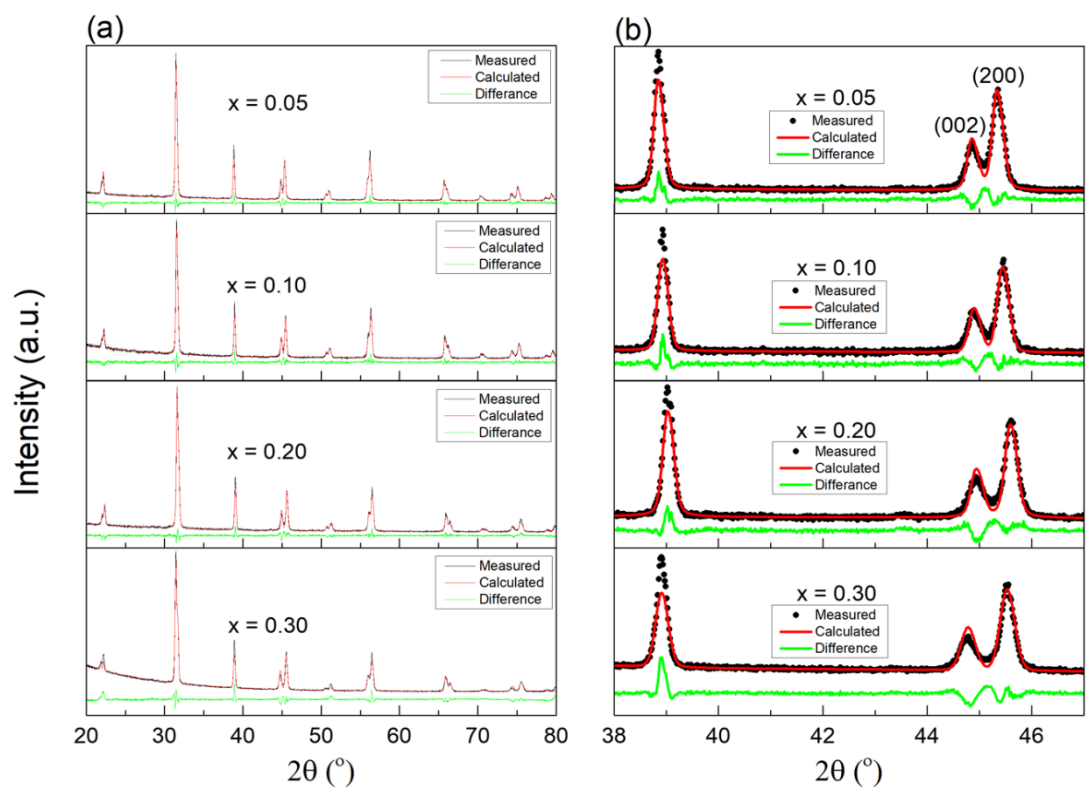

**Fig.1** Refined XRD patterns of (1-x)BaTiO₃-xBi₀.₅Na₀.₅TiO₃ (BT-NBT) ceramics for x = 0.05, 0.10, 0.20 and 0.30 compositions in the range between (a) 20 -80 ° and (b) 38°-47°.

For the direct measurements of the electrocaloric effect, ΔT was measured by a small bead thermistor (GR150KM3976J15, Measurement Specialties, USA) glued to the surface of the sample. The thermistor resistance was measured with a multimeter HP 34401A. Electrical contacts were achieved with thin copper wire glued to the top and bottom sample surface. Voltage to the samples was supplied with Trek 20/20C amplifier, while a step signal was generated with the National

Instruments data acquisition card. A temperature-controlled chamber ensured isothermal conditions with a temperature stabilization of ±0.002 K. Detailed information about the method is given elsewhere.[23]

## Results and discussion

Fig.1(a) shows the x-ray diffraction patterns -including Rietveld fits using the tetragonal P4mm structure of BT - of the BT-NBT ceramics as a function of NBT content (x). No impurity peaks are observed for any compositions. A gradual increase of the tetragonal peak splitting (distinct (002) and (200) reflections) as a function of increasing NBT content can be observed in Fig.1(b). NBT incorporation shifts (200) peak position to higher angles, while relatively little change occurs in the (002) peak position with the increasing NBT composition up until x = 0.20 and then both peaks shift to lower 2θ, going from x = 0.2 to x = 0.3. These shifts correspond to a slight change in the lattice parameter *c* and an obvious decrease in *a* as confirmed by the Rietveld refinement results, in agreement with previous work.[9] In Table 1 and Fig. 2, refinement results are summarized. The systematic increase in *c/a* values shows that the tetragonality is dramatically enhanced by NBT incorporation despite substituting two heterovalent ions (Na[+] and Bi[3+]) at the A-site. Generally, when cations such as Zr, Sn and Sr are doped into $BaTiO_3$, ferroelectric distortion is weakened, and the tetragonality as well as $T_C$ are reduced.[24] The only exception is $Pb^{2+}$ doping at the A-site, which strongly increases the tetragonality and ferroelectric distortion.[25] Here, in contrast to the effect of the other dopants (excluding $Pb^{2+}$), tetragonality is increased by substituting $Na_{0.5}Bi_{0.5}$ into the A-site. The origin of this increase has been explained by Rao et al.[8] They suggest that $Bi^{3+}$, having $6s^2$ lone pair electrons like $Pb^{2+}$, increases the covalent character of the bond between the A-site cation and one oxygen ion, which leads to a large strain and increased tetragonality,[8, 26] in analogy to $PbTiO_3$.

**Table 1** Rietveld refinement results of (1-x)BaTiO₃-xBi₀.₅Na₀.₅TiO₃ (BT-NBT) ceramics.

| SAMPLE | $a$ (Å) | $c$ (Å) | $c/a$ (Å) | Theoretical Density (g/cm$^3$) | wR (%) |
|---|---|---|---|---|---|
| 0.95BT-0.05NBT | 3.9932(2) | 4.0336(2) | 1.0101 | 5.994 | 8.9 |
| 0.90BT-0.10NBT | 3.9886(1) | 4.0346(1) | 1.0115 | 5.979 | 9 |
| 0.80BT-0.20NBT | 3.9778(1) | 4.0327(1) | 1.0137 | 5.958 | 8.7 |
| 0.70BT-0.30NBT | 3.9674(2) | 4.0300(2) | 1.0157 | 5.938 | 6.5 |

The SEM images of the (1-x)BaTiO₃-xBi₀.₅Na₀.₅TiO₃ ceramics sintered at 1200 °C are shown in Figure 3. The average grain sizes of the samples for x = 0.05, 0.10, 0.20 and 0.30 compositions are calculated as 0.56, 0.65, 0.83 and 1.29 μm, respectively, by linear intercept method and their grain size distribution plots are also shown in the insets of Figure 3. The increase in grain size with increasing NBT shows that NBT incorporation enhances grain growth. The relative densities were calculated as 87.7%, 94.7%, 94.4% and 92.5%, for x = 0.05, 0.10, 0.20 and 0.30 compositions, respectively.

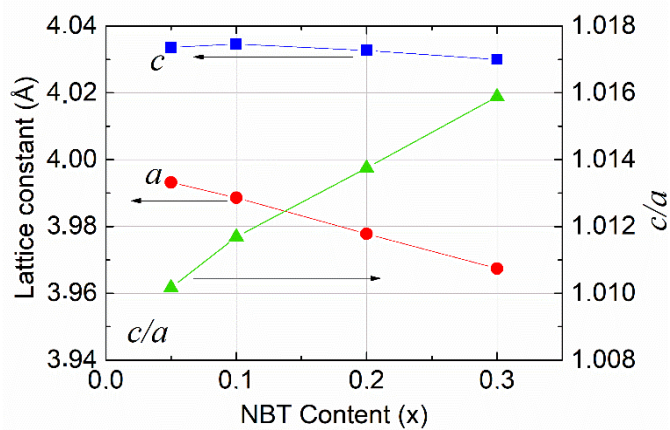

**Fig.2** Lattice parameters and tetragonality (c/a) of BT-NBT ceramics as the function of NBT content according to the Rietveld refinement.

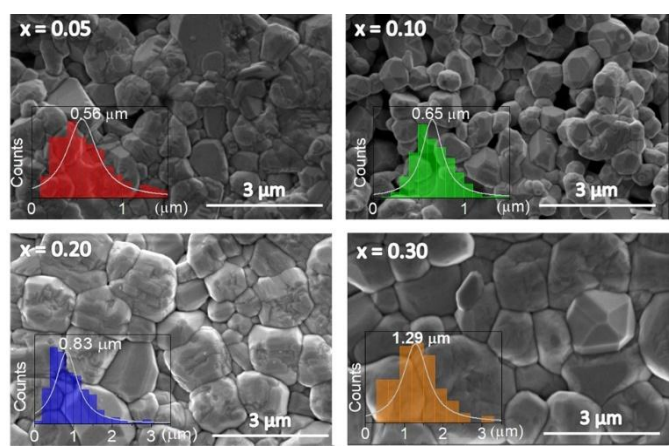

**Fig.3** SEM micrographs and grain size distribution plots of (1-x)BaTiO₃-xNa₀.₅Bi₀.₅TiO₃ ceramics (x = 0.05, 0.10, 0.20 and 0.30).

Fig. 4 shows the temperature-dependent real and imaginary permittivity of the BT-NBT samples. Dielectric peak temperature corresponding to the $T_C$ increases as the NBT content increases, consistent with the previous reports. This is an expected result in accordance with the increase in the tetragonality (*c/a*) with NBT content and stabilization of the ferroelectric phase.[6] Increasing NBT content lowers the dielectric constant at room temperature, which might be due to the decreasing domain wall mobility. The increase in tetragonality can cause a decrease in the domain wall mobility due to the intergranular constraints.[27] On the other hand, the dielectric peak at Tc evolves differently with increasing NBT content. Maximum dielectric constant increases with increasing NBT content, but also becomes broader at the same time. In Fig. S1, temperature dependent dielectric constant and imaginary permittivity of the samples, at four different frequencies of 0.1, 1, 10 and 100 kHz are shown. Dielectric constant is slightly suppressed with increasing frequency. The temperature of the dielectric maxima shifts slightly to higher temperatures only for the x = 0.3 sample. It was reported that it is impossible to observe the higher temperature shift with increasing frequency in the real part of the dielectric constant. Instead, such shift, that occurs in the imaginary part of the dielectric constant was used to claim that the crossover from normal ferroelectric to relaxor ferroelectric behaviour occurs when x = 0.3.[8] Imaginary permittivity of the samples increases with increasing frequency, in agreement with the literature.[8] In order to evaluate the diffuseness level of the phase transition, modified Curie-Weiss law is used: $\frac{1}{\varepsilon} - \frac{1}{\varepsilon_m} = \frac{(T-T_m)^{\gamma}}{C'}$. In this equation, $\varepsilon_m$ corresponds to the maximum dielectric constant, $T_m$ is the temperature at which $\varepsilon_m$ is observed and $\gamma$ and $C'$ are constants. Diffuseness coefficient $\gamma$ for each composition is calculated from the fits to the modified Curie-Weiss law as shown in Figure S2. $\gamma = 1$ describes normal ferroelectric behaviour while $\gamma = 2$ implies relaxor ferroelectric behaviour. According to the fits, gamma coefficient values are 1.22, 1.54, 1.56 and 1.70 for x = 0.05, 0.1, 0.2 and 0.3 samples, respectively. This change implies that as the NBT content increases, the nature of the phase transition changes from normal ferroelectric to diffuse type and eventually approaching relaxor ferroelectric type.

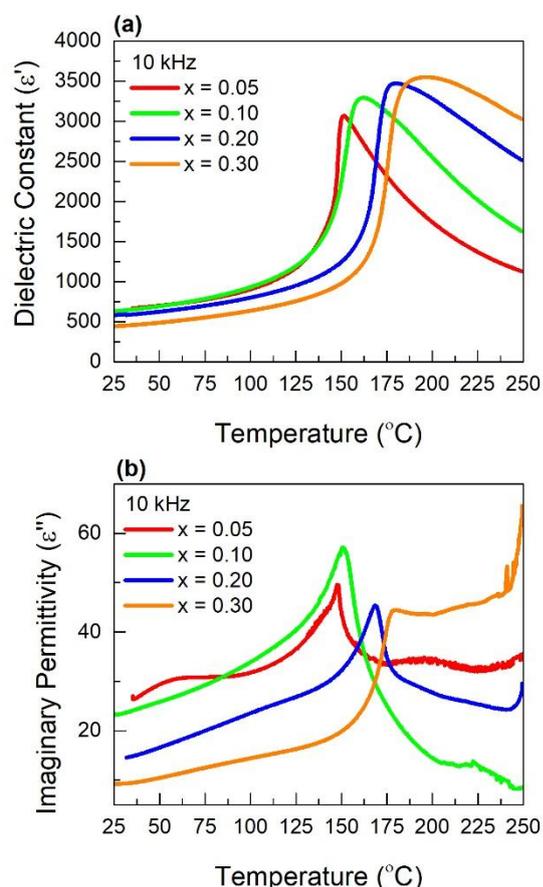

**Fig. 4.** Comparison of (a) dielectric constant and (b) imaginary permittivity of the samples for all compositions at 10 kHz.

In Fig.5, ferroelectric hysteresis loops and strain-electric field curves of all samples at room temperature are collected. It can be observed that both the coercive field and the remanent polarization of the samples increase with increasing NBT content up to x = 0.2. In contrast, x = 0.3 sample has a slightly higher coercive field than x = 0.2 and almost the same $P_r$. These findings are consistent with those of a previous report.[9] We note that the relatively low saturation polarization value for the x = 0.05 sample originates from the low sintering temperature, which was kept deliberately low to avoid the evaporation of volatile components. Electric field-induced strain (S(E)) curves show consistent behaviour with the P(E) measurements. Negative strain is small for the x = 0.05 sample but increases with increasing NBT content, as it is related with the increased remanent polarization of the samples. x = 0.2 and x = 0.3 samples both have large negative strain values.

In Fig. S3, P(E) loops and S(E) curves of all samples are plotted at selected temperatures. As the temperature increases, saturation polarization of the loops initially increases due to the decrease in coercive field because the loops are not fully saturated at room temperature. For example, for x = 0.2, $P_s$ at 95 °C is larger than at room temperature. Further increase in the temperature decreases saturation polarization as expected. As the Curie temperature is approached, for both x = 0.2 and x = 0.3 samples, double hysteresis loops are observed (Fig.S3(e,g)). At even higher temperatures, the loops become almost linear, indicating the transition into the paraelectric phase. The hysteresis loop of the x = 0.05 sample has a slight conductivity contribution, which is evident from a lossy loop around $T_{C} = 150$ °C. For the other samples, no lossy loops were observed, and the P-E loops were slim above $T_c$. In S(E) curves, maximum positive strain values are obtained close to the Curie temperature of each composition. This is unexpected, since for normal ferroelectrics such as BaTiO₃, field-induced strain is expected to decrease in parallel to the decrease in electrical polarization as the temperature is increased towards $T_C$. This kind of increase in strain with increasing temperature is typically observed in NBT-based relaxor ferroelectrics where the electric field-induced transition from the ergodic relaxor to the ferroelectric state induces large strain close to the depolarization temperature.[28]

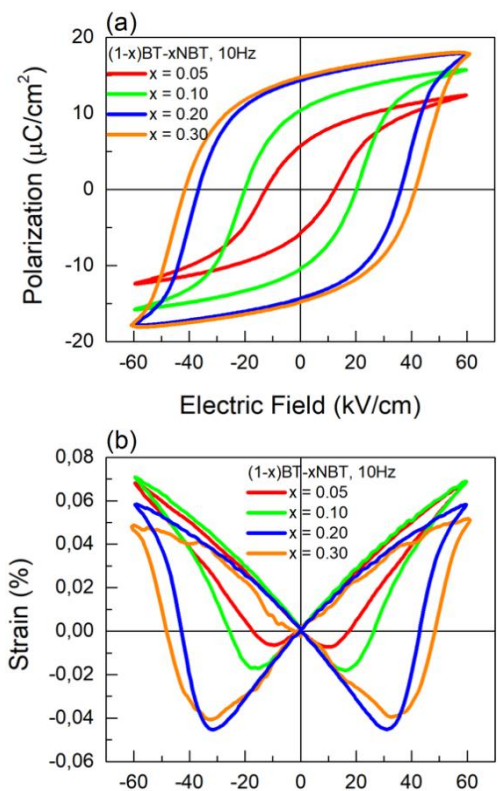

**Fig. 5.** Comparison of P(E) loops and S(E) curves of all samples at 10 Hz.

Temperature dependence of the electrical polarization extracted from the P(E) loops measured at different temperatures and corresponding $\Delta T$ values calculated indirectly at 40 kV/cm are shown in Figure 6. It can be observed that P(T) changes more sharply around $T_C$ for x = 0.2 and x = 0.3 samples, and quite unexpectedly, a sharp $\Delta T$ peak is obtained for these samples. Maximum $\Delta T$ values were obtained as 0.36, 1.30, 2.63, and 3.02 K for compositions x = 0.05, 0.10, 0.20 and 0.30, respectively. We will discuss below that the sharp $\Delta T$ peak at $T_C$ and correspondingly large $\Delta T$ must originate from the unexpected first-order phase transition character of the x = 0.2 and x = 0.3 samples. Typically, by doping (or substitution), the phase transition from the ferroelectric to paraelectric phase broadens, and this causes a decrease in the magnitude and sharpness of the $\Delta T$ peak. Here, quite extraordinarily, despite the broadening of the dielectric peak, order of the phase transition becomes first-order or close to first-order upon increasing the NBT content in the solid solution.

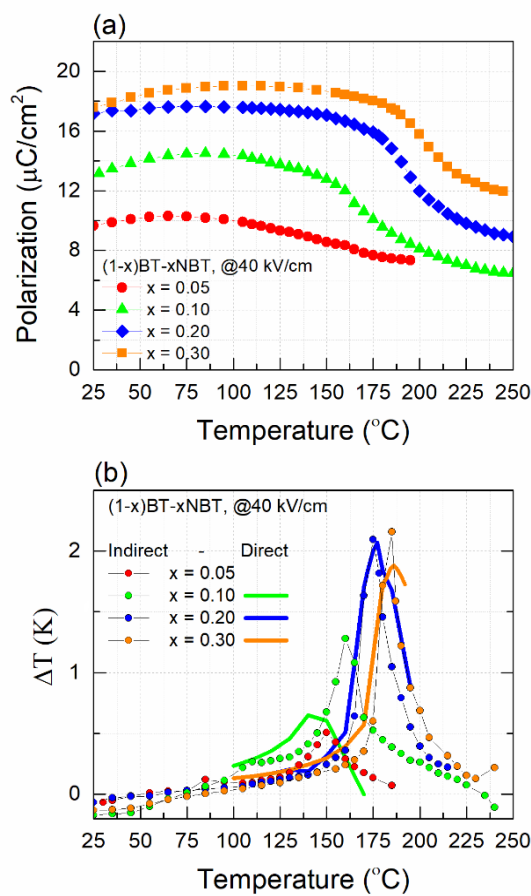

**Fig. 6** (a) P(T) behaviour extracted from the temperature dependent P(E) loops, (b) corresponding $\Delta T$ calculated indirectly and directly measured $\Delta T$, superimposed on the indirect measurement results.

Results of the direct electrocaloric measurements of x = 0.1, 0.2, and 0.3 samples are superimposed on the indirectly calculated $\Delta T$. Direct measurements reveal lower maximum $\Delta T$ values compared to indirect measurements while for x = 0.2 composition, the difference is little. The sample x = 0.2 shows the largest $\Delta T$, slightly over 2 K at 40 kV/cm. The lower $\Delta T$ of the x = 0.3 sample results from the significant Joule heating contribution at temperatures close to $T_C$ during the direct measurement.

Joule heating has been often observed during EC measurements[29, 30] and particularly in Bi-based materials[31]. As shown in the case of $Pb(Fe_{0.5}Nb_{0.5})O_3$-$BiFeO_3$ solid-solution Mn doping can be an approach to reduce Joule heating[31]. However, in the same study, it was verified that the measurement of the EC effect can be performed with certain accuracy despite the presence of Joule heating[31]. The important condition is that Joule heating contribution should not be higher than the contribution from the electrocaloric effect itself. In order to take Joule heating into account and subtract its contribution, we define a term, $\Delta T_D$, representing the total temperature change measured by the direct measurement being equal to the sum of maximum effective EC cooling ($\Delta T_{eff}$) and Joule heating ($\Delta T_{JH}$). A similar approach of decoupling Joule heating from pure electrocaloric effect has been reported for $0.5Ba(Zr_{0.2}Ti_{0.8})O_3$-$0.5(Ba_{0.7}Ca_{0.3})TiO_3$[32].

Direct measurement results, showing $\Delta T_D$ for x = 0.1, 0.2 and 0.3 samples as a function of both temperature and electric field are separately included in Fig. S4. In order to account for the Joule heating, $\Delta T_{eff}$ values are plotted in Fig. S5. In addition, time dependence of $\Delta T$ for x = 0.3 and x = 0.2 samples are shown in Figs. S6 and S7, respectively. On Fig. S6, we show by markings how $\Delta T_{JH}$ is obtained, to be subtracted from $\Delta T_D$. For the three temperatures, close to the Curie temperature for x = 0.3 composition, after the electric field is switched on, $\Delta T$ of the sample immediately rises due to the electrocaloric effect but then relaxes (drops back) due to the thermal diffusion, not to the same temperature but to a higher temperature than the initial temperature due to the Joule heating, represented by $\Delta T_{JH}$. It can be observed from those figures that while x = 0.3 sample suffers from Joule heating, x = 0.2 sample has a minor Joule heating contribution and the $\Delta T_{eff}$ of that sample remains significantly high (~1.65 K).

The discrepancy between direct and indirect measurement results is relatively large in $\Delta T$ as this was measured in two different setups. Furthermore, $\Delta T$ discrepancy between direct and indirect $\Delta T$ has been recognized in the literature[33]. Particularly challenging is obtaining $\left(\frac{dP}{dT}\right)_E$ and later calculations of $\Delta T$ at or near $T_c$. This is mainly due to the discontinuous nature of first-order phase transition. On the other hand, near $T_c$ dielectric losses increase, and thus, the Joule heating during direct measurements affects the measurements. For example, composition with x = 0.3 at 182 to 186 °C exhibits approximately 1 °C of Joule heating. Please note that during direct measurements, the DC electric field is held constant for 50 seconds, while in the case of indirect measurements, P-E loops are measured within 1 second (1 Hz). For applications, more relevant are shorter times, with a frequency range of field application at 0.2-2 Hz; thus, the indirect measurements are of high importance.[34-36]

Electrocaloric strength ($\Delta T_{eff}/\Delta E$) of the x = 0.2 sample amounts to 0.43 K m /MV. This value is slightly lower than the literature results on BT (e.g. 0.60 K m /MV ),[37] despite the increase in the tetragonality and first-order phase transition character by NBT substitution. This might be due to the slightly lower maximum polarization of our samples caused by the NBT substitution.

Experimental results compiled together in Fig. 7 can be used to assess the phase transition character of the samples as a function of NBT content. In Fig. 7(a), DSC measurement results showing the heat flow of the samples as a function of temperature, recorded both on heating and cooling are plotted. Using this data and the data obtained from the measurement of the blank holder and a sapphire reference sample, specific heat of the samples shown in Fig. S8 is obtained. In Fig. 7(a), an apparent thermal hysteresis between heating and cooling can be observed for all samples due to the difference between the peak temperatures at the Curie temperatures. The thermal hysteresis becomes more pronounced with increasing NBT content in the samples, x = 0.3 being the sample with the largest temperature difference between heating and cooling. The thermal hysteresis is also clear in the temperature dependence of the real and imaginary parts of the dielectric constant plotted in Fig. 7(b). Another important observation is the increase in the sharpness of the DSC peaks at the phase transition with NBT content. x = 0.02 sample shows the sharpest peak at the transition, followed by x = 0.03. This does not agree with the thermal hysteresis data, where x = 0.03 shows higher thermal hysteresis than x = 0.02. The temperature dependence of the remanent polarization plotted in Fig. 7(c) shows that the depolarization in x = 0.2 and x = 0.3 samples happens more suddenly than those in the lower NBT-containing samples. In the insets of Fig. 7(c), P(E) loops of all samples at temperatures slightly above the Curie temperature of each sample are shown. It is clear that x = 0.2 and x = 0.3 samples show rare double hysteresis behaviour, while x = 0.05 and 0.10 samples show slightly non-linear and slim loops as expected. Hysteresis loops for all samples at and slightly above Tc are included in Figs. S9-S12, where it can be observed that double loops exist over a wide range of temperatures for the x = 0.2 and 0.3 samples.

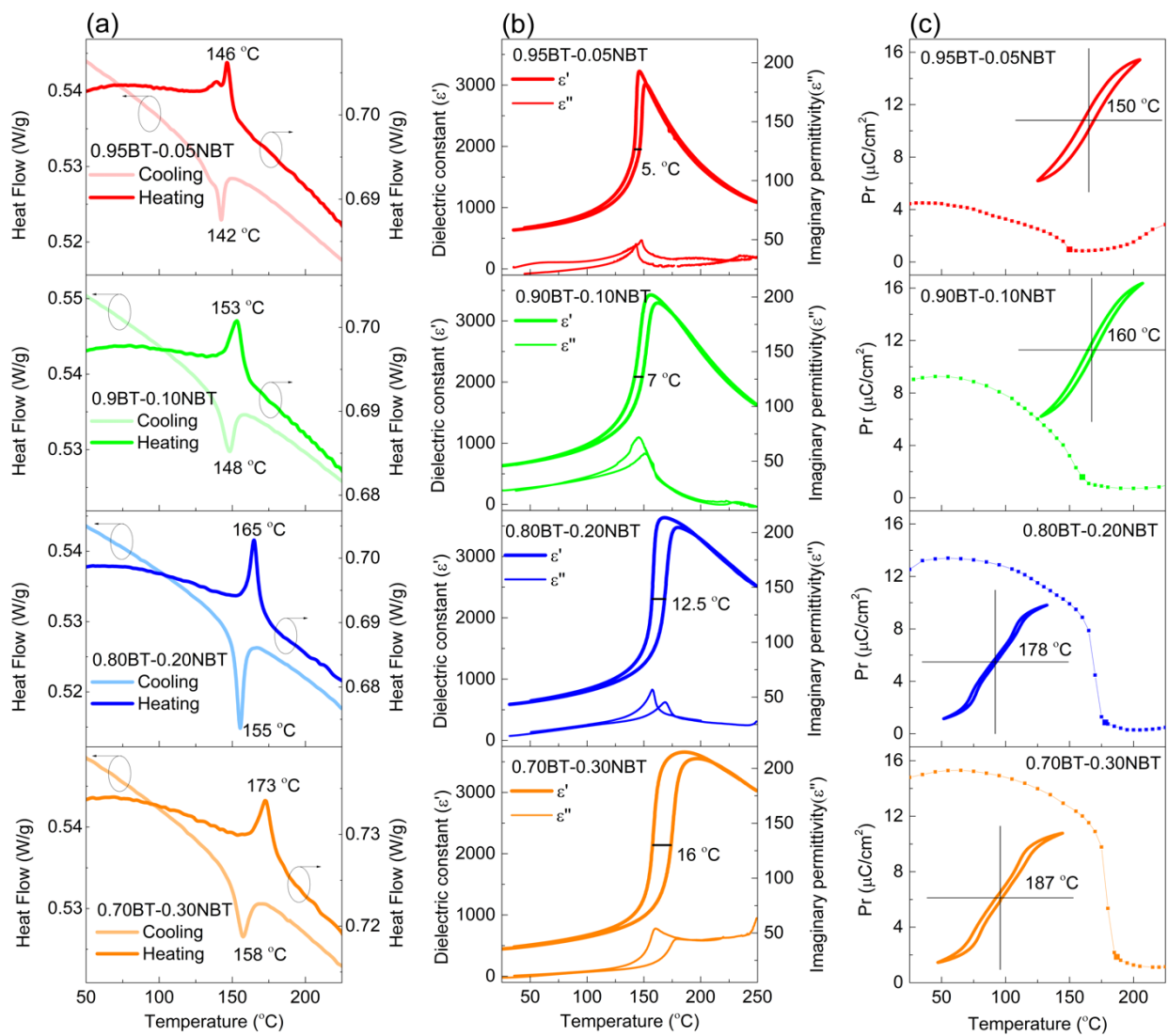

**Fig.7** Comparison of the phase transition behavior of BT-NBT samples. (a) DSC measurements on heating and cooling, (b) dielectric measurements on heating and cooling at 10 kHz, and (c) temperature dependence of remanent polarization. In the insets of the figures in (c), P(E) loops at temperatures close to the Curie temperatures of the respective compositions are shown.

All experimental data in Fig. 7 suggest that NBT-substitution strengthens the ferroelectricity and first-order transition character of the ferroelectric-paraelectric phase transition, despite the compositional and charge disorder it introduces to BT, which is evidenced by the increase in the diffusivity coefficient $\gamma$. Thermal hysteresis in the dielectric constant or heat flow, between heating and cooling is frequently used as an indicator for assessing the nature of the phase transition. If the transition is of the first-order, thermal hysteresis occurs according to the Landau theory.[38] The thermal hysteresis in heat flow and/or dielectric constant was reported for single crystalline $BaTiO_3$,[14] $BaTiO_3$ in multilayer capacitor form,[16] as well as for different ferroelectrics such as a tungsten bronze, and even B-site cation ordered $Pb(Sc_{0.5}Ta_{0.5})O_3$.[19] First-order phase transition was previously claimed by Datta et al. for (1-x)BT-xNBT compositions until x = 0.20 using dielectric measurements.[7] Double-hysteresis loops at and slightly above $T_c$ in ferroelectrics were first reported by Merz on $BaTiO_3$[39] and are later associated with the first-order nature and the sharpness of the ferroelectric-paraelectric phase transition.[15, 40, 41] Electrocaloric effect is therefore expected to be large for normal ferroelectrics with double hysteresis loops, which was already demonstrated in very dense polycrystalline $BaTiO_3$.[15] Therefore, based on our findings in Fig.7 as well as the relatively large $\Delta T_{eff}$ for x = 0.2 sample we measured, we conclude that despite the compositional and charge disorder that it is introduced by NBT, ferroelectric nature of the $BaTiO_3$ is strengthened by NBT substitution, which induces a first-order phase transition and correspondingly large $\Delta T$. Nevertheless, we note that more experiments such as Piezoforce Microscopy (PFM) measurements are necessary to pinpoint the precise crossover composition from the first-order to the second-order phase transition. It is well-known that in the NBT-rich part of the solid solution, double hysteresis loops due to from field-induced transition from ergodic relaxor to ferroelectric state. Domain imaging using PFM can help to identify the state of the samples. As a closing remark, as we discussed in the introduction section, NBT substitution is analogous to $PbTiO_3$ -due to presence of the lone pair ion $Bi^{3+}$-therefore we expect similar results for $Pb^{2+}$ substituted $BaTiO_3$, which is the focus of our following study. We note that around 10 % $PbTiO_3$ substituted $BaTiO_3$ shows a similar tetragonality[25] to our 30 % NBT substituted BT.

## Conclusions

We have studied the electrocaloric response of BT-rich part of NBT-BT solid solution. We have confirmed the increase in the tetragonality and Curie temperature of the samples using XRD and electrical measurements. We have observed a rise in the $\Delta T_D$ of the samples with increasing NBT content. We ascribed this increase to the strengthening of the ferroelectric nature of the samples by increasing NBT substitution, despite the simultaneous increase in the compositional and charge disorder. We show that the first-order nature of the ferroelectric-paraelectric phase transition is also strengthened by using thermal hysteresis in both the dielectric constant and heat flow between heating and cooling, as well as by observing double loops at and slightly above $T_c$. Our results not only present a high temperature electrocaloric material with good electrocaloric response but also highlight the importance of the BT-rich part of the BT-NBT solid solution, which will indeed trigger more fundamental and applied research. In addition, for the electrocaloric cooling applications, it should be possible to decrease the Curie temperature close to room temperature by appropriate doping, without compromising too much from ΔT.

## Conflicts of interest

There are no conflicts of interest to declare.

## Data Availability

The data supporting this article have been included as part of the Supplementary Information.

## Acknowledgements

We acknowledge IZTECH Centre for Materials Research for the use of XRD, SEM and DSC instruments. We thank M. Barış Okatan for his assistance with the indirect calculation of the electrocaloric effect and useful discussions and Oğuz Akkaşoğlu for useful discussions.

SUPPLEMENTARY INFORMATION

for

# Stabilization of the first-order phase transition character and Enhancement of the Electrocaloric Effect by NBT substitution in BaTiO₃ ceramics

**Merve Karakaya[a], İrem Gürbüz[a,b], Lovro Fulanovic[c], Umut Adem[a]**

[a]Department of Materials Science and Engineering, İzmir Institute of Technology, Urla, 35430, İzmir Turkey

[b]Department of Chemistry, Aix-Marseille University, 52 Avenue Escadrille Normandie Niemen, 13013, Marseille, France.

[c]Nonmetallic Inorganic Materials, Department of Materials and Earth Sciences, Technical University of Darmstadt, Peter-Grünberg-Straße 2, 64287 Darmstadt, Germany

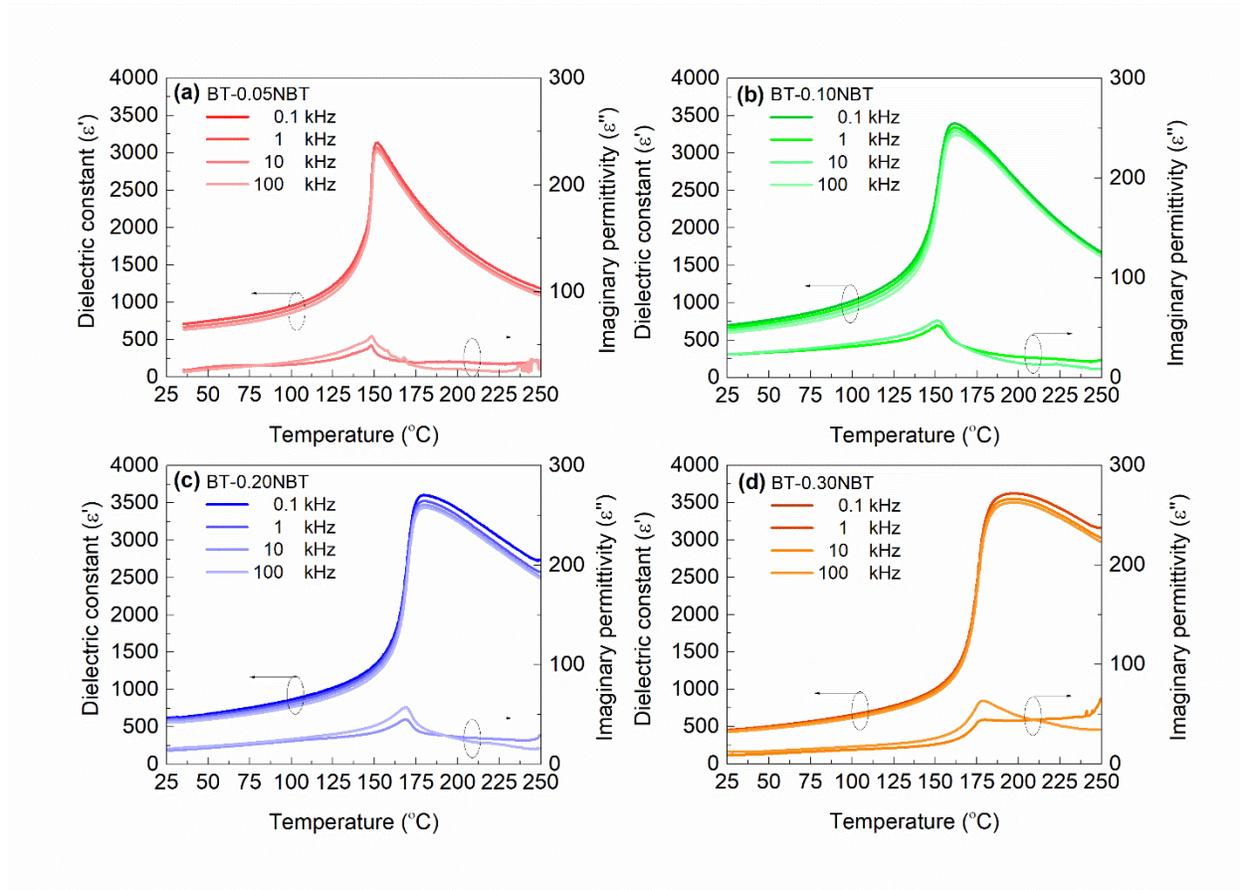

**Fig. S1.** Temperature dependence of the dielectric constant and imaginary permittivity at 0.1, 1, 10 and 100 kHz of (1-x)BaTiO$_3$-xNa$_{0.5}$Bi$_{0.5}$TiO$_3$ samples: (a) x = 0.05, (b) x = 0.10, (c) x = 0.20, (d) x = 0.30.

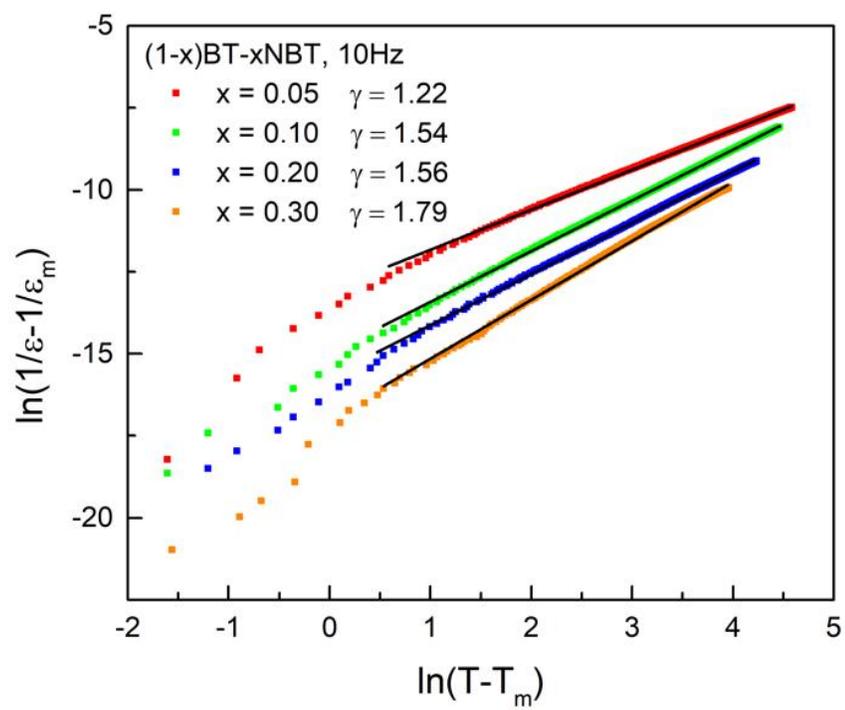

**Fig. S2.** The plot of $\ln(1/\varepsilon - 1/\varepsilon_m)$ as a function of $\ln(T-T_m)$ for all compositions, showing $\gamma$ coefficients and linear fits.

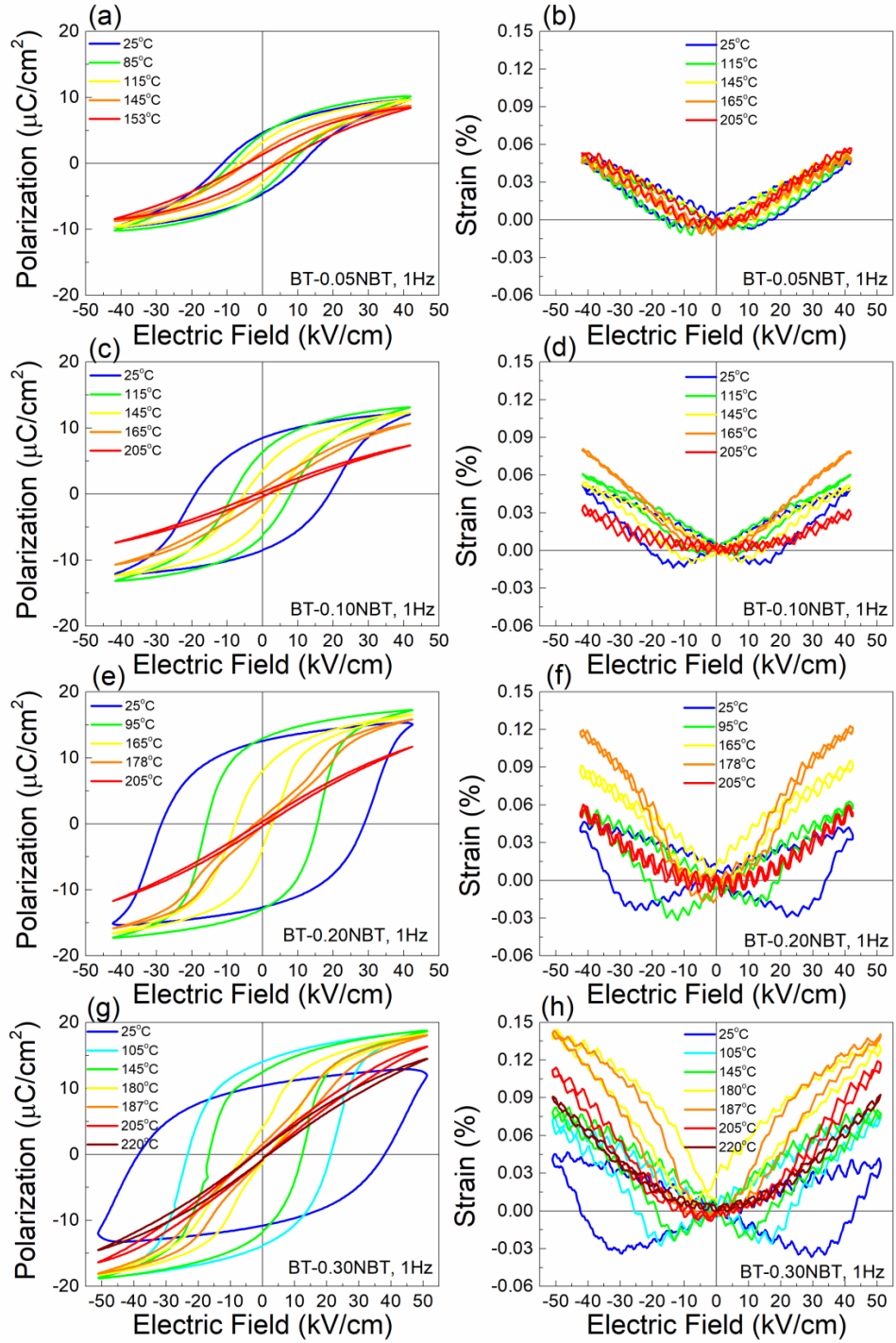

**Fig. S3.** Temperature dependent P(E) hysteresis loops and strain-electric field curves at selected temperatures. (a, b) x = 0.05, (c, d) x = 0.10, (e, f) x = 0.20, (g, h) x = 0.30.

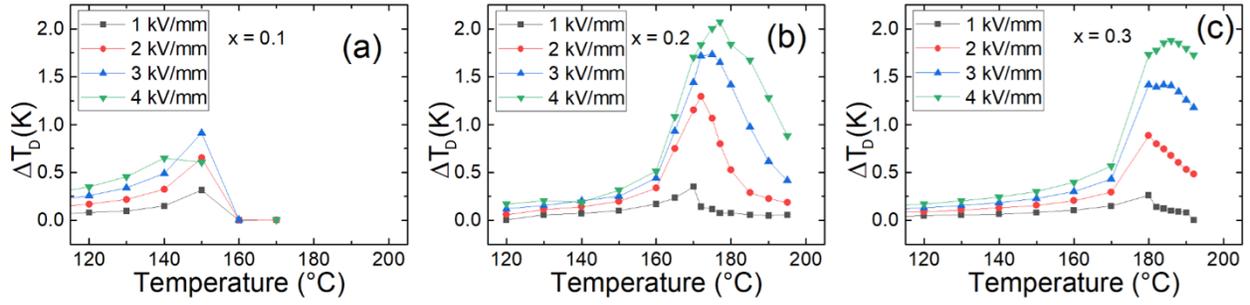

**Fig. S4.** Temperature and electric field dependence of directly measured $\Delta T_D$ for (a) x = 0.1, (b) x = 0.2 and (c) x = 0.3 samples.

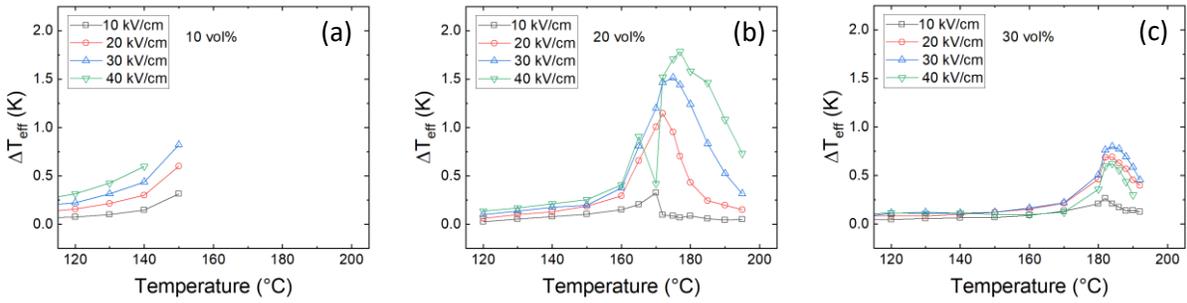

**Fig. S5.** Temperature and electric field dependence of directly measured effective EC cooling $\Delta T_{eff}$ for (a) x = 0.1, (b) x = 0.2 and (c) x =0.3 samples. $\Delta T_{eff} = \Delta T_D - \Delta T_{JH}$

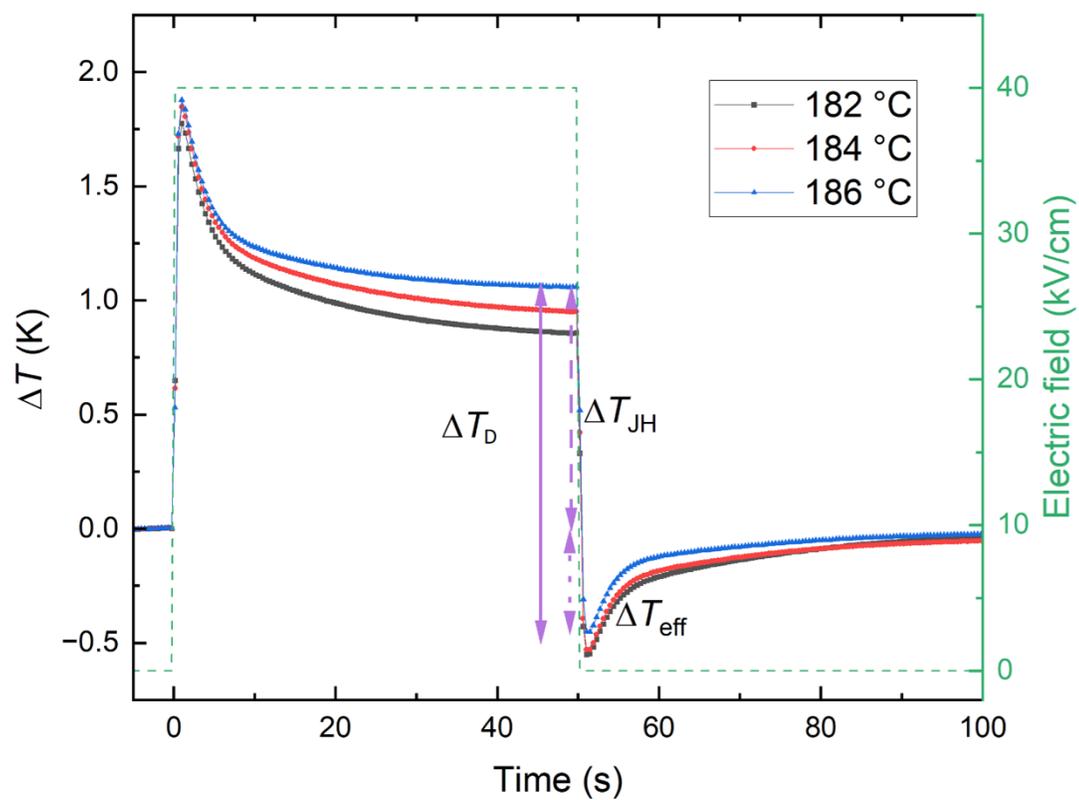

**Fig. S6.** Directly measured $\Delta T$ for 0.7BT-0.3NBT sample close to the dielectric maximum temperature as a function of time as the electric field was turned on and off. $\Delta T_D$ is the sum of the maximum effective EC cooling ($\Delta T_{eff}$) and the Joule heating ($\Delta T_{JH}$).

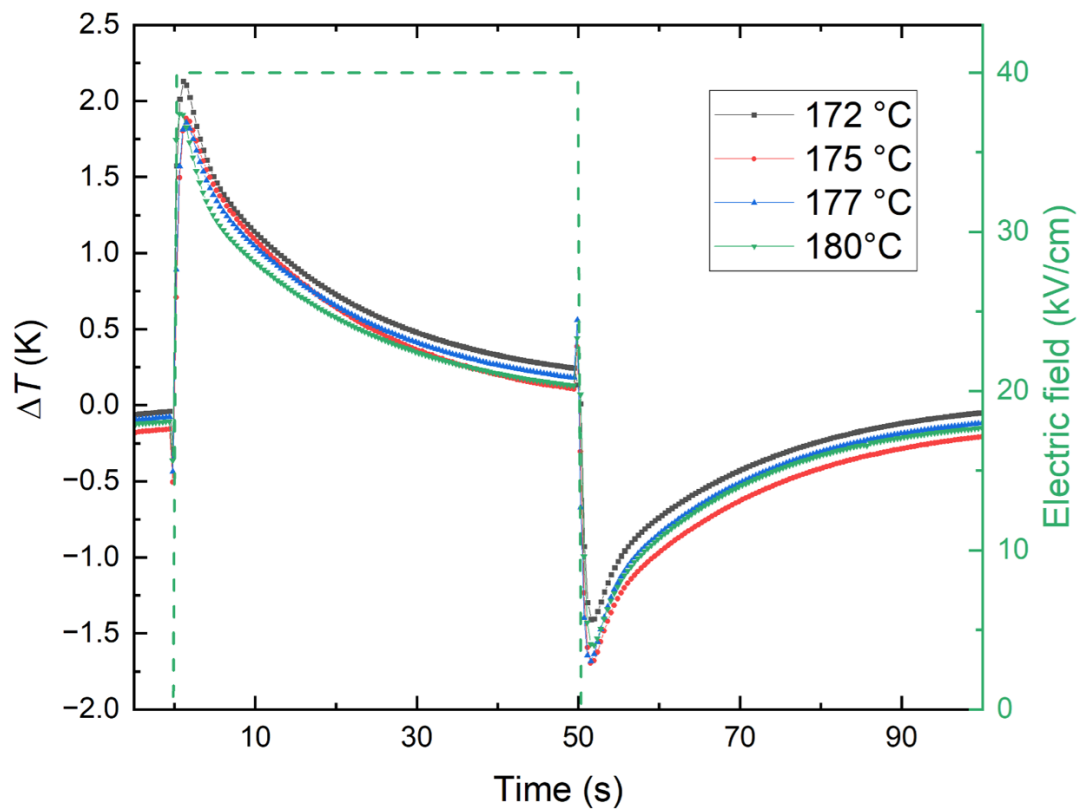

**Fig. S7.** Directly measured *ΔT* for 0.8BT-0.2NBT sample close to the dielectric maximum temperature as a function of time as the electric field was turned on and off.

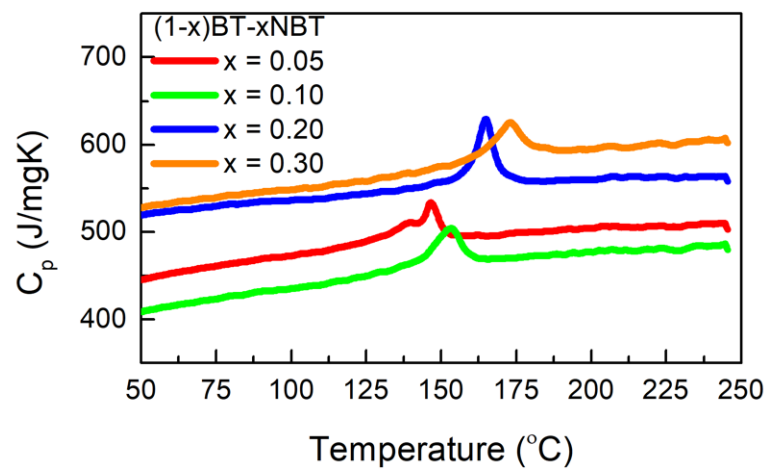

**Fig. S8.** Temperature dependence of specific heat of (1-x)BaTiO₃-xNa₀.₅Bi₀.₅TiO₃ samples.

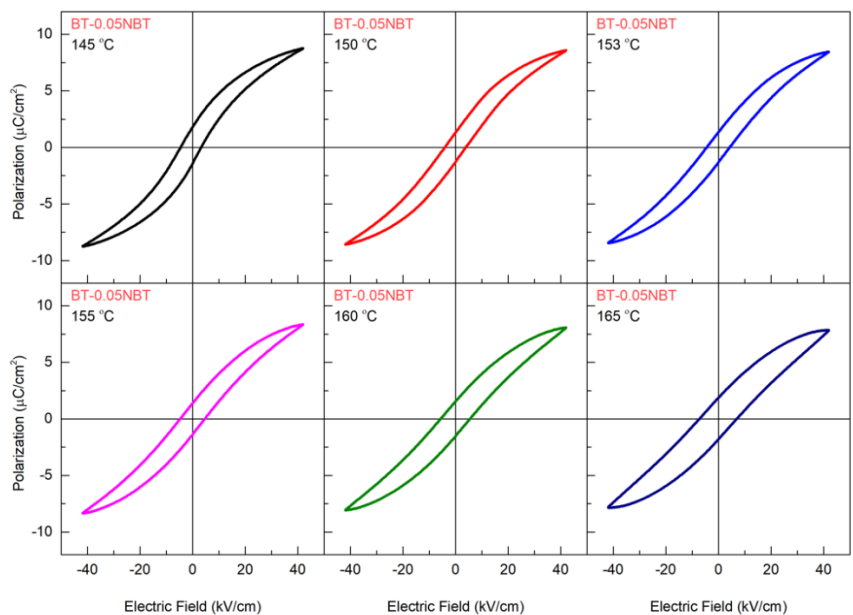

**Fig. S9.** P(E) hysteresis loops of 0.95BT-0.05NBT at and slightly above T$_C$.

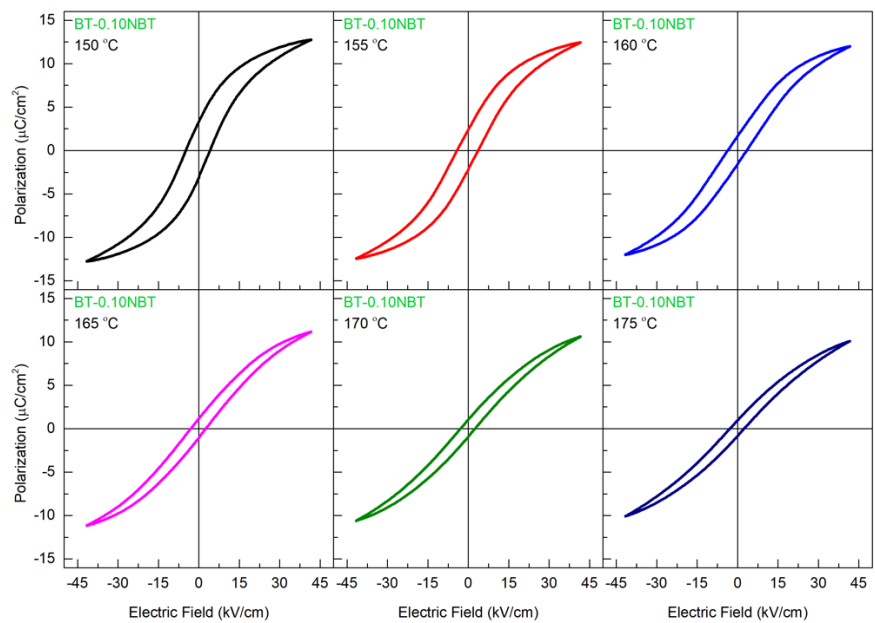

**Fig. S10.** P(E) hysteresis loops of 0.9BT-0.1NBT at and slightly above T$_C$.

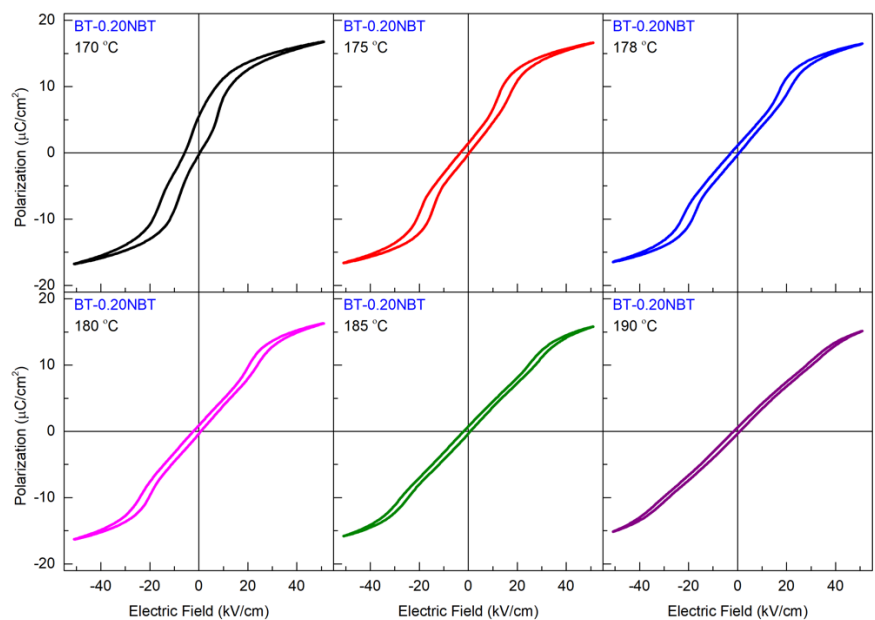

**Fig. S11.** P(E) hysteresis loops of 0.8BT-0.2NBT at and slightly above T$_C$.

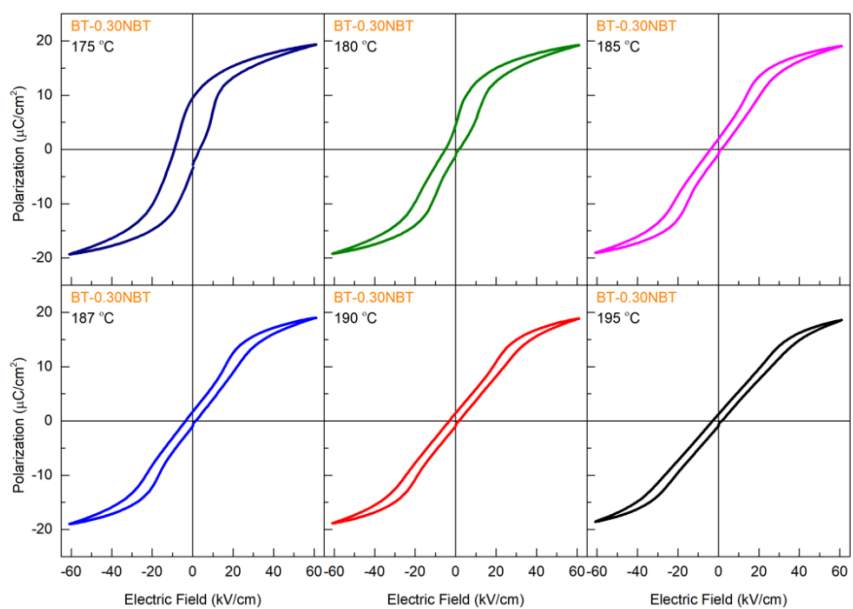

**Fig. S12.** P(E) hysteresis loops of 0.7BT-0.3NBT at and slightly above T$_C$.